\documentclass[amsmath,amssymb,numbers]{revtex4}
\begin{document}

\title{Maximal Entanglement via Collective Coordinates}

\author{M. Revzen}
\affiliation {Department of Physics, Technion - Israel Institute of Technology, Haifa
32000, Israel}

\date{\today}

\begin{abstract}
Maximal entangled states (MES) provide a basis to 2d-dimensional particles Hilbert space, d=prime $\ne2$. These states allow generalization of the Mean King Problem. The states may be viewed as build of points each underpins a product state carrying a mutual unbiased bases (MUB) label or, alternatively, as product states labeled with center of mass and relative coordinates. The coordinate-like label of the center of mass and the momentum-like of the relative coordinates provides a MES account of the Hilbert space in close analogy with the single particle phase space coordinates.    
\end{abstract}

\pacs{03.65.Ta;03.65.Wj;02.10.Ox}

\maketitle

\section {  Introduction}

Entanglement is fundamental to quantum mechanics \cite{schroed}. It plays a central role in virtually   all analyses involving the definitive attributes of quantum theory. In this work we show that maximal entanglement may be accounted for via basically classic notion: Independence of center of mass and relative coordinates of the constituent particles system \cite{rev1}. thus we consider two d-dimensional particles Hilbert space, $d=prime\;\ne2$. The system may always be accounted for via collective, center of mass and relative coordinates. A product state in the collective coordinates is a MES. We consider, aided with Fivel's analysis of MES \cite{fivel}, the class of such states that are product states in the collective coordinates basis and show that they provide a MES orthonormal basis spanning the two d-dimensional particles Hilbert space.\\
Viewing the maximally entangled state as a product state of the collective variables is then shown to have a particular geometrical meaning which allow an intuitive interpretation to the solution of the Mean King Problem \cite{durt, lev, berge} as well its generalization to the Tracking of the Mean King, \cite{rev2}. The geometrical view allows \cite{rev3} a convenient Wigner function like mapping of {\it one} d-dimensional particle Hilbert space onto c-number functions of lines and points of the underpinning geometry and, in particular, to formulate a finite dimensional Radon transformation \cite{rev3}. The present study gives intuitive extension of the geometrical analysis of the single particles study to two particles systems.  In so doing we identify what is termed \cite{rev2} a balancing term of geometrical origin that allows the inversion of interrelations among the Hilbert space operators/states.\\
Brief review on several topics are included in an attempt to have the paper self contained. Following 
Weyl \cite{weyl} and Schwinger \cite{schwinger} we use unitary operators to represent physical states.\\

\section{Brief Review: Mutually Unbiased Bases (MUB) and Mutually Unbiased Collective Bases (MUCB)}

In a d-dimensional Hilbert space two complete, orthonormal vectorial bases, ${\cal
B}_1,\;{\cal B}_2$,
 are said to be MUB if and only if (${\cal B}_1\ne {\cal B}_2)$
\begin{equation}
\forall |u\rangle,\;|v \rangle\; \epsilon \;{\cal B}_1,\;{\cal B}_2 \;resp.,\;\;|\langle
u|v\rangle|=1/\sqrt{d}.
\end{equation}
Maximal number of MUB allowed in a d-dimensional Hilbert space is d+1 \cite{ivanovich,bengtsson}. Variety of methods for construction of the d+1 bases for $d=p^m$ are now available
\cite{wootters2,tal,klimov2,vourdas}. Our present study is confined to $d=p\;\ne2$.\\
 We now give explicitly the MUB states in conjunction with the algebraically complete
 operators \cite{schwinger,amir} set:
 $\hat{Z},\hat{X}$.  Thus we label the d distinct states spanning the Hilbert space,
 termed
 the computational basis (CB), by $|n\rangle,\;\;n=0,1,..d-1; |n+d\rangle=|n\rangle$

\begin{equation}
\hat{Z}|n\rangle=\omega^{n}|n\rangle;\;\hat{X}|n\rangle=|n+1\rangle,\;\omega=e^{i2\pi/d}.
\end{equation}
Aside from the CB the d states in each of the d MUB bases are \cite{tal,amir} 
\begin{equation} \label{mxel}
|m;b\rangle=\frac{1}{\sqrt
d}\sum_0^{d-1}\omega^{\frac{b}{2}n(n-1)-nm}|n\rangle;\;\;b,m=0,1,..d-1.
\end{equation}
Here the d sets labeled by b are the bases and the m labels the states within a basis. Each basis
relate to a unitary operator, \cite{tal}, $\hat{X}\hat{Z}^b|m;b\rangle=\omega^m|m;b\rangle.$

For later reference we shall refer to the computational basis (CB) by $b=\ddot{0}$. Thus the above
gives d+1 bases, $b=\ddot{0},,0,1,...d-1$ with the total number of states d(d+1) grouped in d+1 sets
each of d states.\\
The MUB set is closed under complex conjugation,
\begin{equation}\label{cc}
\langle n|m,b\rangle^{\ast}=\langle n|\tilde{m},\tilde{b}\rangle,\;\Rightarrow|\tilde{m},\tilde{b}\rangle=|d-m,d-b\rangle,
\end{equation}
as can be verified from Eq.(\ref{mxel}).\\

Several studies \cite{durt,klimov1,fivel,berge,rev1} consider the entanglement of two
d-dimensional particles Hilbert space via MUB state labeling. We shall now outline briefly
the approach adopted by \cite{rev1} that will be used in later sections.\\
The Hilbert space is spanned by the single particle computational bases,
$|n\rangle_1|n'\rangle_2$ (the subscripts denote the particles). These are eigenfunctions
of $\hat{Z}_i$ i=1,2:
$\hat{Z}_i|n\rangle_i=\omega^{n}|n\rangle_i,\;\omega=e^{i\frac{2\pi}{d}}.$ Similarly
$\hat{X}_i|n\rangle_i=|n+1\rangle,\;i=1,2$. We now define our collective coordinates and
collective operators (we remind the reader that the exponents are modular variables, e.g.
1/2 mod[d=7]=(d+1)/2)=4):
\begin{equation}\label{colz}
\hat{Z}_r\equiv \hat{Z}^{1/2}_{1}\hat{Z}^{-1/2}_{2};\;\;\bar{Z}_c\equiv
\hat{Z}^{1/2}_{1}\hat{Z}^{1/2}_{2}\;\leftrightarrow\;\hat{Z}_1=\hat{Z}_r\hat{Z}_c;\;\;\hat{Z}_2=\hat{Z}_r^{-1}\hat{Z}_c,
\end{equation}
and, in a similar manner,
\begin{equation}\label{colx}
\hat{X}_r\equiv\hat{X}_1\hat{X}_2^{-1};\;\hat{X}_c\equiv\hat{X}_1\hat{X}_2\rightarrow
\hat{X}_1=\hat{X}^{1/2}_r\hat{X}^{1/2}_c,\;\hat{X}_2=\hat{X}^{-1/2}_r\hat{X}^{1/2}_c.
\end{equation}

Since $\bar{Z}_{s}^{d}=\bar{X}_s^{d}=1,$  and   $\bar{X}_s\bar{Z}_s=\omega\bar{Z}_s\bar{X}_s,\;s=r,c;\;\bar{X}_s\bar{Z}_{s'}=\bar{Z}_{s'}\bar{X}_s,\;s\ne
s',$ we may consider their respective computational
eigen-bases and with it the whole set of MUB bases, \cite{rev1}, 
\begin{equation}
\bar{Z}_{s}|n\rangle_s = \omega^{n}|n\rangle_s,\;\; \bar{X}_s\bar{Z}_s^{b_s}|m_s,b_s\rangle=\omega^{m_s}|m_s,b_s\rangle;\;\;\langle n_s|m_s,b_s\rangle=\omega^{\frac{b_s}{2}n_s(n_s-1)-m_sn_s}.\;s=r,c,
\end{equation}
clearly  $|n\rangle_r|n'\rangle_c;\;n,n'=0,1,..d-1,$ is a $d^2$ orthonormal basis spanning
the two d-dimensional particles Hilbert space.
One readily proves \cite{rev1}
\begin{equation}\label{relcomdel}
\langle n_1,n_2|n_r,n_c\rangle=\delta_{n_r,(n_1-n_2)/2}\delta_{n_c,(n_1+n_2)/2}.
\end{equation}
We have then,
\begin{equation}\label{relcom}
|n_r,n_c\rangle=|n_1,n_2\rangle,\;\;for\;n_r=(n_1-n_2)/2,\;n_c=(n_1+n_2)/2\;\rightleftarrows
n_1=n_r+n_c,\;n_2=n_c-n_r.
\end{equation}
There are, of course, d+1 MUB bases for each of the collective modes. We adopt here too the notational simplification $b_s\rightarrow \ddot{0}_s,\; s=r,c$.

\section{   Finite Geometry and Hilbert Space Operators, Brief Outline}

We now briefly review the essential features of finite geometry required for our study
\cite{bennett,wootters4,shirakova,tomer,saniga,rev2,rev3}.\\
A finite plane geometry is a system possessing a finite number of points and lines. there are two kinds of finite plane geometry affine and projective. We confine ourselves to affine plane geometry (APG).\\ 
It can be shown \cite{bennett,shirakova} that for $d=p^m$ (a power of prime) APG can be
constructed (our study here is for d=p).  Furthermore the existence of APG implied the existence of its 
dual geometry DAPG wherein the points and lines are interchanged. Since we shall study extensively DAPG we list its properties \cite{bennett, shirakova}. We shall refer to these by DAPG(.)\\

a. The number of lines is $d^2$, $L_j,\;j=1,2....d^2.$ The number of points is d(d+1),
$S_{\alpha},\;{\alpha = 1,2,...d(d+1)}.$\\
b. A pair of points on a line determine a line uniquely. Two (distinct) lines share one and only
one point.\\ 
c. Each point is common to d lines. Each line contain d+1 points: $S_{\alpha}=\bigcap_{j\in \alpha}^d L_j;\;\;L_j=\bigcup_{\alpha \in j}^{d+1}S_{\alpha}.$\\
d. The d(d+1) points may be grouped in sets of d points no two of a set
share a line. Such a set is designated by $\alpha' \in \{\alpha \cup M_{\alpha}\},\;
\alpha'=1,2,...d$. ($M_{\alpha}$ contain all the points not connected to $\alpha$ - they
are not connected among themselves.) i.e. such a set contain d disjoint (among themselves)
points. There are d+1 such sets:
\begin{equation}
\bigcup_{\alpha=1}^{d(d+1)}S_{\alpha}=\bigcup_{\alpha=1}^d R_{\alpha};\;\;
R_{\alpha}=\bigcup_{\alpha'\epsilon\alpha\cup M_{\alpha}}S_{\alpha'};\;\;
R_{\alpha}\bigcap R_{\alpha'}=\varnothing,\;\alpha\ne\alpha'.
\end{equation}
e. Each point of a set of disjoint points is connected to every other point not in its
set.\\
DAPG(c) allows the transcription, which we adopt, of $S_{\alpha}$ in terms of {\it addition} of $L_j$.
This acquires a meaning upon viewing the points ($S_{\alpha}$) and the lines ($L_j$) as underpinning Hilbert space entities, e.g. projectors or states, to be specified later. This point is further discussed at the end of this section:
\begin{equation} \label{sum}
S_{\alpha}=\frac{1}{d}\sum_{j \in \alpha}^{d}L_j\;\;\Rightarrow\;\;\sum_{\alpha{'}\in \alpha\cup M_{\alpha}}^d S_{\alpha{'}}=\frac{1}{d}\sum_j^{d^2} L_j.
\end{equation}
DAPG(d) via Eq.(\ref{sum}) implies
\begin{equation} \label{R}
{\cal{R}}\equiv \sum_{\alpha{'}\in \alpha\cup M_{\alpha}}^d S_{\alpha{'}}=\frac{1}{d}\sum_j^{d^2} L_j=
\frac{1}{d+1}\sum_{\alpha}^{d(d+1)} S_{\alpha}\;\;independent\;of\;\alpha.
\end{equation}
This equation, Eq.(\ref{R}), reflects relation among equivalent classes within the geometry \cite{bennett}. It will be referred to as the balance formula: ${\cal{R}}$ serves as a balancing term.
Thus Eqs.(\ref{sum}),(\ref{R}) imply:
\begin{equation}\label{line}
L_j\;=\;\sum_{\alpha \in j}^{d+1} S_{\alpha}\;-\sum_{\alpha{'}\in \alpha\cup M_{\alpha}}^d S_{\alpha
{'}}\;=\;\sum_{\alpha \in j}^{d+1} S_{\alpha}\;-\;{\cal{R}}.
\end{equation}

A particular arrangement of lines and points that satisfies APG(x), x=a,b,c,d,e is referred to as a realization of  DAPG.\\

We now consider a particular realization of DAPG of dimensionality $d=p,\ne 2$ which is the
basis of our present study. We arrange the aggregate the d(d+1) points, $\alpha$, in a
$d\cdot(d+1)$matrix like rectangular array of d rows and d+1 columns. Each column is made
of a set of d points  $R_{\alpha}=\bigcup_{\alpha'\epsilon\alpha\cup
M_{\alpha}}S_{\alpha'};$  DAPG(d). We label the columns by b=-1,0,1,2,....,d-1 and the rows
by m=0,1,2...d-1.( Note that the first column label of $b=\ddot{0}$ is for convenience with no numerical implication.)   $\alpha=m(b)$ designate a point by its row, m,
and its column, b; when b is allowed to vary - it designate the point's row position in
every column. We label the left most column by $b=\ddot{0}$ and with increasing values of b, the
basis label, we move to the right. Thus the right most column is b=d-1.\\
In a realization of DAPG via Hilbert space operators or states, letting A stand for the Hilbert space entity underpinned with the coordinated point, (m,b), in \cite{rev3} A stood for a projector,
$A_{\alpha=(m,b)}\;\rightarrow\;\hat{A}_{\alpha}=|m,b\rangle\langle b,m|.$ In the present work A will signify two particles product state to be specified in a subsequent section. We now assert
that the d+1 points, $m_j(b), b=\ddot{0},0,1,2,...d-1,$  form the line j which
contain the two (specific) points $(m,\ddot{0})$ denoted by $\ddot{m}$ and $(m,0)$ denoted $m_0$ is given by (we forfeit the subscript j - it is implicit - $j=(\ddot{m},m_0)$),
\begin{equation} \label{m(b)}
m(b)=\;m_0\;+\frac{b}{2}(2\ddot{m}-1),\;\;b\ne \ddot{0};\;\;m(\ddot{0})=\ddot{m}.
\end{equation}
For the underpinning considered in \cite{rev2, rev3}, viz $S_{\alpha}\;\rightarrow\;\hat{A}_{\alpha}$ the balance term is ${\cal{R}}=\sum_{m=0}^{d-1}|m,b\rangle\langle b,m|=\Bbb{I}$, i.e. independent of $\alpha,j$ and consistent with the balancing equation requirement, Eq.(\ref{R}). In \cite{rev2,rev3}
the resultant line, the sum of projectors,  is an operator, $L_j\;\rightarrow\;\hat{P}_j$ Eq.(\ref{line}) becomes $\hat{P}_j=\sum_{\alpha \in j}\hat{A}_{\alpha}-\Bbb{I}$ which is shown \cite{rev2,rev3} to abide by
\begin{equation}\label{P}
\langle n|\hat{P}_j|n'\rangle=\delta_{n+n',2\ddot{m}}\omega^{-(n-n')m_0}\;\;\hat{P}_j^2=\Bbb{I};\;\forall\;j\;
\;tr\hat{P}_j\hat{P}_{j'}=d\delta_{j,j'}.
\end{equation}

Our present case with $A_{\alpha=(m,b)}\;\rightarrow\;|A_{\alpha=(m,b)}\rangle$ will lead to a universal ($\alpha$ 
independent) vector, $|{\cal{R}}\rangle$, defined in the next section.\\
It is possible to consider different transcription corresponding to DAPG(c), \cite{tomer}, e.g. instead of Eq.(\ref{sum}) one may take,
$$L_j\;=\;\frac{1}{d+1}\sum_{\alpha \in j}^{d+1}S_{\alpha}.$$
This transcription, though consistent, leads to more complicated formalism.\\

\section{Geometric Underpinning of Two Particles States}

We now consider DAPG underpinning for {\it states} of a {\it two} d-dimensional Hilbert space. The coordination scheme is as considered above, $\alpha=(m,b);\;j=(\ddot{0},m_0).$ The line equation is given by Eq.(\ref{m(b)}). However the point $S_{\alpha}$ underpins the state (the numerical subscripts refers to the particles)
\begin{equation}\label{2p}
|A_{\alpha=(m,b)}\rangle\;=\;|m,b\rangle_1|\tilde{m},\tilde{b}\rangle_2.
\end{equation}
$|\tilde{m},\tilde{b}\rangle$ is given by Eq.(\ref{cc}). Eqs.(\ref{sum},\ref{line}) now read,
\begin{equation}
|A_{\alpha}\rangle=\frac{1}{d}\sum_{j\in\alpha}^d|P_j\rangle\;\;\Rightarrow\;\;|P_j\rangle=\sum_{\alpha \in j}^{d+1}|A_{\alpha}\rangle\;-\;|{\cal{R}}\rangle;\;\;|{\cal{R}}\rangle=\sum_{\alpha{'} \in \alpha\cup M_{\alpha}}|A_{\alpha{'}}\rangle.
\end{equation}
(Note that the "line" state, $|P_j\rangle$ is not normalized.) We now utilize our choice, Eq.(\ref{cc}), to show the 
universality (independence of the basis, b,  in this case) of ${\cal{R}}\rangle$,\cite{durt,fivel}
\begin{equation}\label{uniR}
|{\cal{R}}\rangle=\sum_{\alpha{'}\in \alpha \cup M_{\alpha}}|A_{\alpha}\rangle=
\sum_{m\in b}|m,b\rangle|\tilde{m},\tilde{b}\rangle=\sum_{m,n,n'}|n\rangle_1|n'\rangle_2 \langle n|m,b\rangle \langle n'|\tilde{m},\tilde{b}\rangle=\sum_{n}|n\rangle_1|n\rangle_2, \;\;indep.\; of\; b.
\end{equation}
The relation among the matrix elements of the projectors, $\hat{A}_{(m,b)}=|m,b\rangle\langle b,m|$,
residing on the line, Eq.(\ref{m(b)}) \cite{rev2,rev3} and the two particle states $|A_{(m,b)}\rangle=|m,b\rangle|\tilde{m},\tilde{b}\rangle$, residing on the equivalent line, Eq.(\ref{m(b)}), are now used to obtain an explicit expression for the "line" state,$|P_j\rangle$,
\begin{eqnarray}\label{line2}
|P_{j=(\ddot{m},m_0)}\;&=&\;\frac{1}{\sqrt d}\sum_{m(b)\in j}|m,b\rangle_1|\tilde{m},\tilde{b}\rangle_2
-|{\cal{R}}\rangle= \nonumber \\
\frac{1}{\sqrt d}\sum_{n,n'}|n\rangle_1|n'\rangle_2\big[\langle n|\sum_{m(b)\in j}\hat{A}_{(m,b)}-\Bbb{I}|n'\rangle\big]\;&=&\;\frac{1}{\sqrt d}\sum_{n,n'}|n\rangle_1|n'\rangle_2
\delta_{n+n',2\ddot{m}}\omega^{-(n-n')m_0},\;\forall b.
\end{eqnarray}
We used Eqs. (\ref{cc}), (\ref{P}) and (\ref{uniR}). This formula will now be put in a more pliable form \cite{fivel} and then expressed in terms of the collective coordinates,
\begin{eqnarray}\label{list}
|P_{j=(\ddot{m},m_0)}\;&=&\;\frac{1}{\sqrt d}\sum_{n,n'}|n\rangle_1|n'\rangle_2\delta_{n+n',2\ddot{m}}\omega^{-(n-n')m_0}=\nonumber \\ \frac{\omega^{2\ddot{m}m_0}}{\sqrt d}\sum_n|n\rangle_1|2\ddot{m}-n\rangle_2\omega^{-2nm_0}&=& 
\frac{\omega^{2\ddot{m}m_0}}{\sqrt d}\sum_n|n\rangle_1\hat{X}^{2\ddot{m}}\hat{Z}^{2m_0}{\cal{I}}\tau|n\rangle_2 =\nonumber \\
\frac{\omega^{2\ddot{m}m_0}}{\sqrt d}\sum_m|m,b\rangle_1\hat{X}^{2\ddot{m}}\hat{Z}^{2m_0}{\cal{I}}|\tilde{m},\tilde{b}\rangle_2&=&
|\ddot{m}\rangle_c|2m_0\rangle_r.
\end{eqnarray}
The inversion operator ${\cal{I}}$ is defined via ${\cal{I}}|n\rangle=|-n\rangle$, and we used our definition of $\tau$ Eq.(\ref{cc}) :$\tau|m,b\rangle=|\tilde{m},\tilde{b}\rangle;\;\tau|n\rangle=|n\rangle.$ The last equality in Eq.(\ref{list}), follows upon noting that
\begin{equation} 
|\ddot{m}\rangle_c|2m_0\rangle_r=\frac{1}{\sqrt d}|\ddot{m}\rangle_c\sum_n|n\rangle_r\omega^{-2m_0n}=\frac{1}{\sqrt d}\sum_n|\ddot{m}+n\rangle_1|\ddot{m}-n\rangle_2\omega^{-2m_0n}.
\end{equation}
That the $d^2$ vectors $|P_{j=\ddot{m},m_0}\rangle$ form an orthonormal set is manifest in the collective coordinates formulation.\\
The central result of our geometrical approach is the following intuitively obvious overlap relation,
\begin{eqnarray}\label{cent}
\langle A_{\alpha=(m,b)}|P_{j=(\ddot{m},m_0)}\rangle\equiv \langle m,b|_1\langle\tilde{m},\tilde{b}|_2P_{j=\ddot{m},m_0}\rangle\;&=&\;\frac{1}{\sqrt d}\delta_{m,(m_0+\frac{b}{2}[2\ddot{m}-1])},\;b\ne\ddot{0},\nonumber \\
\langle A_{\alpha=(n,\ddot{0})}|P_{j=(\ddot{m},m_0)}\rangle\equiv\langle n|_1\langle n|_2P_{j=\ddot{m},m_0}\rangle\;&=&\;\frac{1}{\sqrt d}\delta_{n,\ddot{m}},\;\;b=\ddot{0}.
\end{eqnarray}
Thus the overlap of $|A_{\alpha=(m,b)}\rangle$ with $|P_j\rangle$ vanishes for $\alpha\not\in j$, i.e.
for $m\ne m_0+b/2(2\ddot{m}-1)$. i.e. only if the point (m,b) is on the line j the overlap is not zero.
This, as shall see shortly, is the key to the solution of the MKP \cite{berge}. This is a remarkable attribute as it holds for the particle {\it pair} while each of the constituent by itself does not abide by it. We further note that the probability of finding the state $|A_{\alpha}\rangle$ given that the state of the system is $|P_j\rangle$ with $\alpha\in j$ is 1/d while the the number of points ($\alpha$) on the line is d+1 exposing these probabilities to be not mutually exclusive.\\

\section{Leaky particle}
The maximally entangled state, Eq.(\ref{list}), was viewed as a "line" state, i.e. as sum of product states each underpinned by the geometrical point, $\alpha=(m,b)$, Eq.(\ref{2p}). The coordinates, (m,b), abide by the line equation, Eq.(\ref{m(b)}), form a line in the sense that, cf. Eq.(\ref{cent}),
\begin{equation}
\langle A_{\alpha}|P_j\rangle=\langle \tilde{b},\tilde{m}|_2\langle b,m|_1\ddot{m}\rangle_c|2m_0\rangle_r=\begin{cases}\frac{1}{d},\;\alpha\in j,\\ 0,\;\alpha \not\in j.
\end{cases}
\end{equation}  
Thus particle {\it pair}, viz. the particle and its mate, the tilde particle, whose coordinates are 
$\alpha=m,b$ do as a whole belong to the d lines that share this coordinate. However each of of the constituent particles (either 1 or 2) is {\it equally} likely to be in any of the $d^2$ lines:
\begin{equation}\label{single}
\langle b,m|_1\ddot{m}\rangle_c|2m_0\rangle_r=\frac{1}{\sqrt d}|\tilde{\bar{m}}+\tilde{\Delta},\tilde{b}\rangle_2\omega^{-2\ddot{m}\Delta},\;
\bar{m}=m_0+\frac{b}{2}(2\ddot{m}-1);\;\;\Delta=\bar{m}-m.
\end{equation}
 Thence,
\begin{equation}
|\langle b,m|_1\ddot{m}\rangle_c|2m_0\rangle_r|^2=\frac{1}{d},\;\forall\;\ddot{m},m_0.
\end{equation}
It is this attribute that allows the tracking of the King's measurement alignment.

\section{Tracking the Mean King}

The Mean King Problem (MKP) was initiated by \cite{lev}, and was analyzed in several publications,
 a comprehensive list is given in \cite{durt}. Briefly summarized it runs as follows. Alice may prepare a state to her liking. The King measures it in some MUB basis (i.e. for some value of b, his choice for the alignment of his apparatus). He does not inform Alice of his observational result nor the basis he used. Alice now performs a control measurement of her liking. {\it After} her control measurement the King informs her the basis (b) he used for his measurement. Thence she {\it deduces} the actual outcome ( the m value) that he observed. This , from our approach is readily done: The state prepared by Alice is ${\cal{R}}\rangle$, Eq.(\ref{uniR}). The King measures in the alignment b the operator,
\begin{equation}\label{K}
\hat{K}_b=\sum_{m=0}^d|m,b\rangle\omega_m\langle b,m|
\end{equation}     
and observes, say $\Omega_m$. The King's measurement projects the state ${\cal{R}}\rangle$ to the state $|A_{\alpha=m,b}\rangle$, Eq.(\ref{2p}).\\
Now for her control measurement Alice measures the non degenerate operator,
\begin{equation}\label{A}
\hat{B}\equiv\sum_{m',m"}|\ddot{m}{'}\rangle_c|2m_0^{"}\rangle_r\Gamma_{m',m"}\langle 2m_0^{"}|_r\langle\ddot{m}{'}|_c,
\end{equation}
obtaining, say, $\Gamma_{m',m"}$. Thence the quantity,
\begin{equation}
\langle A_{\alpha=m,b}|P_{j=\ddot{m}{'},m_0{"}}\rangle\;\ne\;0,
\end{equation} 
i.e. (cf. Eq.(\ref{cent})) $m=m_0{"}+(b/2)(2\ddot{m}{'}-1).$ Since she knows $\ddot{m}{'}$ and $m_0{"}$, through her control measurement, upon being informed of b she infers m, the King's measurement outcome, \cite{lev,berge}. This suggest the following geometrical view. The King's measurement, Eq.(\ref{K}), of the state prepared by Alice, ${\cal{R}}\rangle$, leaves the system at a geometrical point corresponding to Eq.(\ref{2p}). It, cf. DAPG(c), is shared by d lines. Alice's control measurement relate a line, $|P_{j=\ddot{m},m_0}\rangle$, to this point, Eq.(\ref{cent}). The line's equation gives m, the vectorial coordinate, as a function of b, the basis alignment. Thus upon being told the value of b Alice can deduce the outcome via m(b).\\
In our case of {\it tracking} the King - he does not inform Alice the basis he used - her control measurement is designed to track it. To this end the state that Alice prepares is one of the line vectors $|P_{j=\ddot{m},m_0}\rangle$, Eq.(\ref{list}). Thus she knows $\ddot{m}$ and $m_0$. The King's measurement is as in the MKP case, Eq.(\ref{K}), and, as specified above, he observed, $\Omega_m$. In this case the King's measurement projects the state prepared by Alice, $|P_{j=\ddot{m},m_0}\rangle$, to (neglecting normalization) to the much richer state, 
\begin{equation}
|m,b\rangle_1\langle b,m|_1\ddot{m}\rangle_c|m_0\rangle_r.
\end{equation}
Now Alice in her control measurement measures, much like in the MKP, the operator $\hat{B}$,  Eq.{\ref{A}) and obtains, say, like in the case above, $\Gamma_{m',m"}$, implying that
\begin{equation}
\langle 2m_0{"}|_r\langle\ddot{m}{'}|_cm,b\rangle_1\langle b,m|_1\ddot{m}\rangle_c|m_0\rangle_r\;\ne\;0.
\end{equation}
The LHS of this relation is obtained via Eq.(\ref{single}) to equal, up to a  phase,
\begin{equation} 
\langle 2m_0{"}|_r\langle\ddot{m}{'}|_cm,b\rangle_1\langle b,m|_1\ddot{m}\rangle_c|m_0\rangle_r=\frac{1}{d}\delta_{(m_0{"}-m_0),b(\ddot{m}-\ddot{m}{'})},\;b\ne\ddot{0}.
\end{equation}
Thus,
\begin{equation}
b=\frac{(m_0{"}-m_0)}{(\ddot{m}-\ddot{m}{'})};\;\;\ddot{m}=\ddot{m}{'}\rightarrow b=\ddot{0}.
\end{equation}
Knowing the initial state, i.e. $\ddot{m}$, $m_0$, and measuring (in the control measurement) 
$\ddot{m}{'}$ and $m_0{"}$ Alice tracks the King's apparatus alignment, b. The alignment $b=\ddot{0}$, the King use of the CB, gives $\delta_{\ddot{m},\ddot{m}{"}}$. The case wherein both $\ddot{m}{'}=\ddot{m}$ {\it and} $m_0=m_0{"}$ is undetermined.\\

\section{Summary and Concluding Remarks}

Two d-dimensional particles Hilbert space was spanned by maximally entangled states (MES). These MES were associated with lines of finite geometry and the two points that determine a line were shown to have a simple collective coordinates meaning: one gives the center of mass and the other the relative coordinates of the particles. The coordination of the geometry is based on our study of single particle mutual unbiased bases (MUB) labels, columns relates to bases (designated by b) and rows to vectors within a basis (designated by m). The lines run long varying MUB, i.e. it is m as a function of b, m(b), emulating thereby paths in the continuum studies by Svetlichny, \cite{svetlichny}. Indeed the line underpinning a MES is labeled by classical like phase space coordinates which, however, in the present study relates to  collective variables.\\
The association of MES with lines allows an intuitive view of the so called Mean King problem \cite{berge}. We consider another retrodiction predicament: Alice prepares an initial state that would allow her, upon performing her control measurement, to determine the King's apparatus alignment ( b value). This case of tracking the King's measurement requires the initial state to be a MES underpinned with a whole geometrical line. The King measurement, in tis case, does not leave the system in a well defined geometrical point but rather in a more complex, yet pure, state. Alice control measurement gives directly the King's apparatus alignment used in measuring the initial state.\\
The lines and points of finite geometry were shown \cite{rev3} to provide a convenient underpinning for {\it one} d-dimensional particle's Hilbert space. In that case they provided a simple intuitive mapping of the d dimensional Hilbert space {\it operators} onto c number function of the geometry's lines and points in close analogy with the role of phase space for Wigner function in the continuum. For the two d- dimensional particles considered in the present work the geometry provides a convenient underpinning for MES bases that span the $d^2$ dimensional Hilbert space. These MES through their relation to geometry and their account via collective coordinates were shown to allow a novel view of retrodiction in quantum theory.\\

 Acknowledgments: The hospitality of the Perimeter Institute where this work was completed and discussions with Prof. D. Gottesman, L. Hardy and M. Mueller are gratefully acknowledged.\\

\end{document}